%% FM 00-2; mp_arc \# 00-24
%**start of header
\newcount\mgnf\newcount\tipi\newcount\tipoformule
\newcount\aux\newcount\piepagina\newcount\xdata
\mgnf=0
\aux=1           %1 produce aux
\tipoformule=0   %0 usa aux; 1 no (usa i simboli dati)
\piepagina=1     %0 =data e #par.#pag; 1=data e #pag; 2=#pag
\xdata=0         %0 data del giorno, 1 data fissa da \Di:
\def\Di{}

\ifnum\mgnf=1 \aux=0 \tipoformule =1 \piepagina=1 \xdata=1\fi
\newcount\bibl
%\bibl= ?               % 0= rif [XXX], 1= rif. numerici
\ifnum\mgnf=0\bibl=0\else\bibl=1\fi
\bibl=0

% Per poter cambiare a piacimento il formato dei riferimenti
% bibliografici in <nome>.tex:
%
% 1: citare nella forma esemplificata da \ref{B}{2}{20}}
%    ove XXX e' un simbolo per le iniziali e 2 distingue i lavori con
%    le stesse iniziali,  7 e' il numero SIMBOLICO del riferimento per XXX2.
%    Il numero 7 puo' essere ARBITRARIO e viene automaticamente
%    riaggiustato al momento della compilazione (vedi punto 4)
%
% 2: Se si sceglie \bibl=0 si cita nella forma [XXX2]; se si sceglie
%    \bibl=1 si cita nella forma [numero di ordine di prima citazione].
%
% 3: La bibliografia va scritta nella forma \def{\qqq}{<citazione>}
%    in ordine alfabetico per autore attribuendo un simbolo
%    qualsiasi al testo <citazione} (ad es \def\REFCIT{citazione} e
%    va seguita dal comando  \rif{XXX}{2}{\qqq}{0}
%    scritto all' inizio di una riga altrimenti bianca. L' ultimo {0}
%    e' un parametro inutile ed e' li solo provvisoriamente per memoria
%    (di tentativi in corso).
%
% 4: Per il riaggiustamento automatico (in ordine di citazione)
%    occorre compilare (ottenendo oltre ai soliti la scheda ausiliaria
%    ref.b) e poi si deve eseguire il
%    programma rif <nome> che (usando ref.b) produce fin.tex e <nome>.tex
%    con i riferimenti giusti
%    in \bibl=1 e la si ricompila e stampa. La scheda iniziale <nome>.tex
%    diventa <nome>.old.
\ifnum\bibl=0
\def\ref#1#2#3{[#1#2]\write8{#1@#2}}
\def\rif#1#2#3#4{\item{[#1#2]} #3}
\fi

\ifnum\bibl=1
\openout8=ref.b
\def\ref#1#2#3{[#3]\write8{#1@#2}}
\def\rif#1#2#3#4{}

\fi

\def\9#1{\ifnum\aux=1#1\else\relax\fi}
\ifnum\piepagina=0 \footline={\rlap{\hbox{\copy200}\
$\st[\number\pageno]$}\hss\tenrm \foglio\hss}\fi \ifnum\piepagina=1
\footline={\rlap{\hbox{\copy200}} \hss\tenrm \folio\hss}\fi
\ifnum\piepagina=2\footline{\hss\tenrm\folio\hss}\fi

\ifnum\mgnf=0 \magnification=\magstep0
\hsize=13.5truecm\vsize=22.5truecm \parindent=4.pt\fi
\ifnum\mgnf=1 \magnification=\magstep1
\hsize=16.0truecm\vsize=22.5truecm\baselineskip14pt\vglue5.0truecm
\overfullrule=0pt \parindent=4.pt\fi

\let\a=\alpha \let\g=\gamma \let\d=\delta
 \let\z=\zeta \let\h=\eta
\let\th=\vartheta\let\k=\kappa \let\l=\lambda \let\m=\mu \let\n=\nu
\let\x=\xi \let\p=\pi \let\r=\rho \let\s=\sigma \let\t=\tau
 \let\f=\varphi\let\ch=\chi \let\ps=\psi 
  \let\D=\Delta 
\let\X=\Xi   
  
{\count255=\time\divide\count255 by 60 \xdef\oramin{\number\count255}
\multiply\count255 by-60\advance\count255 by\time
\xdef\oramin{\oramin:\ifnum\count255<10 0\fi\the\count255}}
\def\ora{\oramin }

%\Di e' definito all' inizio e ridefinito qui
\ifnum\xdata=0
\def\data{\number\day/\ifcase\month\or gennaio \or
febbraio \or marzo \or aprile \or maggio \or giugno \or luglio \or
agosto \or settembre \or ottobre \or novembre \or dicembre
\fi/\number\year;\ \ora}
\def\Di{\number\day\kern2mm\ifcase\month\or gennaio \or
febbraio \or marzo \or aprile \or maggio \or giugno \or luglio \or
agosto \or settembre \or ottobre \or novembre \or dicembre
\fi\kern0.1mm\number\year}
\else
\def\data{\Di}
\fi

\setbox200\hbox{$\scriptscriptstyle \data $}
\newcount\pgn \pgn=1
\def\foglio{\number\numsec:\number\pgn
\global\advance\pgn by 1} \def\foglioa{A\number\numsec:\number\pgn
\global\advance\pgn by 1}
\global\newcount\numsec\global\newcount\numfor \global\newcount\numfig
\gdef\profonditastruttura{\dp\strutbox}
\def\senondefinito#1{\expandafter\ifx\csname#1\endcsname\relax}
\def\SIA #1,#2,#3 {\senondefinito{#1#2} \expandafter\xdef\csname
#1#2\endcsname{#3} \else \write16{???? ma #1,#2 e' gia' stato definito
!!!!} \fi} \def\etichetta(#1){(\veroparagrafo.\veraformula) \SIA
e,#1,(\veroparagrafo.\veraformula) \global\advance\numfor by 1
\9{\write15{\string\FU (#1){\equ(#1)}}} \9{ \write16{ EQ \equ(#1) == #1
}}} \def \FU(#1)#2{\SIA fu,#1,#2 }
\def\etichettaa(#1){(A\veroparagrafo.\veraformula) \SIA
e,#1,(A\veroparagrafo.\veraformula) \global\advance\numfor by 1
\9{\write15{\string\FU (#1){\equ(#1)}}} \9{ \write16{ EQ \equ(#1) == #1
}}} \def\getichetta(#1){Fig.  \verafigura \SIA e,#1,{\verafigura}
\global\advance\numfig by 1 \9{\write15{\string\FU (#1){\equ(#1)}}} \9{
\write16{ Fig.  \equ(#1) ha simbolo #1 }}} \newdimen\gwidth \def\BOZZA{
\def\alato(##1){ {\vtop to \profonditastruttura{\baselineskip
\profonditastruttura\vss
\rlap{\kern-\hsize\kern-1.2truecm{$\scriptstyle##1$}}}}}
\def\galato(##1){ \gwidth=\hsize \divide\gwidth by 2 {\vtop to
\profonditastruttura{\baselineskip \profonditastruttura\vss
\rlap{\kern-\gwidth\kern-1.2truecm{$\scriptstyle##1$}}}}} }
\def\alato(#1){} \def\galato(#1){}
\def\veroparagrafo{\number\numsec}\def\veraformula{\number\numfor}
\def\verafigura{\number\numfig}
\def\geq(#1){\getichetta(#1)\galato(#1)}
\def\Eq(#1){\eqno{\etichetta(#1)\alato(#1)}}
\def\eq(#1){\etichetta(#1)\alato(#1)}
\def\Eqa(#1){\eqno{\etichettaa(#1)\alato(#1)}}
\def\eqa(#1){\etichettaa(#1)\alato(#1)}
\def\eqv(#1){\senondefinito{fu#1}$\clubsuit$#1\write16{No translation
for #1} \else\csname fu#1\endcsname\fi}
\def\equ(#1){\senondefinito{e#1}\eqv(#1)\else\csname e#1\endcsname\fi}
\openin13=#1.aux \ifeof13 \relax \else \input #1.aux \closein13\fi
\openin14=\jobname.aux \ifeof14 \relax \else \input \jobname.aux
\closein14 \fi \9{\openout15=\jobname.aux} \newskip\ttglue

%\font\dodicirm=cmr12\font\dodicibf=cmbx12\font\dodiciit=cmti12
%\font\titolo=cmbx12 scaled \magstep2
%\font\ottorm=cmr8\font\ottoi=cmmi8\font\ottosy=cmsy8
%\font\ottobf=cmbx8\font\ottott=cmtt8\font\ottosl=cmsl8\font\ottoit=cmti8
%\font\sixrm=cmr6\font\sixbf=cmbx6\font\sixi=cmmi6\font\sixsy=cmsy6

\font\titolo=cmbx10 scaled \magstep1
\font\ottorm=cmr8\font\ottoi=cmmi7\font\ottosy=cmsy7
\font\ottobf=cmbx7\font\ottott=cmtt8\font\ottosl=cmsl8\font\ottoit=cmti7
\font\sixrm=cmr6\font\sixbf=cmbx7\font\sixi=cmmi7\font\sixsy=cmsy7

\font\fiverm=cmr5\font\fivesy=cmsy5\font\fivei=cmmi5\font\fivebf=cmbx5
\def\ottopunti{\def\rm{\fam0\ottorm}\textfont0=\ottorm%
\scriptfont0=\sixrm\scriptscriptfont0=\fiverm\textfont1=\ottoi%
\scriptfont1=\sixi\scriptscriptfont1=\fivei\textfont2=\ottosy%
\scriptfont2=\sixsy\scriptscriptfont2=\fivesy\textfont3=\tenex%
\scriptfont3=\tenex\scriptscriptfont3=\tenex\textfont\itfam=\ottoit%
\def\it{\fam\itfam\ottoit}\textfont\slfam=\ottosl%
\def\sl{\fam\slfam\ottosl}\textfont\ttfam=\ottott%
\def\tt{\fam\ttfam\ottott}\textfont\bffam=\ottobf%
\scriptfont\bffam=\sixbf\scriptscriptfont\bffam=\fivebf%
\def\bf{\fam\bffam\ottobf}\tt\ttglue=.5em plus.25em minus.15em%
\setbox\strutbox=\hbox{\vrule height7pt depth2pt width0pt}%
\normalbaselineskip=9pt\let\sc=\sixrm\normalbaselines\rm}

\catcode`@=11
\def\footnote#1{\edef\@sf{\spacefactor\the\spacefactor}#1\@sf
\insert\footins\bgroup\ottopunti\interlinepenalty100\let\par=\endgraf
\leftskip=0pt \rightskip=0pt \splittopskip=10pt plus 1pt minus 1pt
\floatingpenalty=20000
\smallskip\item{#1}\bgroup\strut\aftergroup\@foot\let\next}
\skip\footins=12pt plus 2pt minus 4pt\dimen\footins=30pc\catcode`@=12
\let\nota=\ottopunti

%% Grafica
\newdimen\xshift \newdimen\xwidth \newdimen\yshift

\def\ins#1#2#3{\vbox to0pt{\kern-#2 \hbox{\kern#1
#3}\vss}\nointerlineskip}

\def\eqfig#1#2#3#4#5{ \par\xwidth=#1
\xshift=\hsize \advance\xshift by-\xwidth \divide\xshift by 2
\yshift=#2 \divide\yshift by 2 \line{\hglue\xshift \vbox to #2{\vfil #3
\includegraphics{#4.ps} }\hfill\raise\yshift\hbox{#5}}} 

\def\8{\write13}

%%%%%%%%%%%%%%%%%%%%%%

\def\V#1{{\,\underline#1\,}}
\def\T#1{#1\kern-4pt\lower9pt\hbox{$\widetilde{}$}\kern4pt{}}
\let\dpr=\partial\def\Dpr{{\V\dpr}} \let\io=\infty\let\ig=\int
\def\fra#1#2{{#1\over#2}}\def\media#1{\langle{#1}\rangle}\let\0=\noindent
\def\guida{\leaders\hbox to 1em{\hss.\hss}\hfill}
\def\tende#1{\ \vtop{\ialign{##\crcr\rightarrowfill\crcr
\noalign{\kern-1pt\nointerlineskip} \hglue3.pt${\scriptstyle%
#1}$\hglue3.pt\crcr}}\,} \def\otto{\
{\kern-1.truept\leftarrow\kern-5.truept\to\kern-1.truept}\ }

\def\pagina{\vfill\eject}\let\ciao=\bye

\def\st{\scriptscriptstyle}
\def\*{\vskip0.3truecm}

\def\lis#1{{\overline #1}}\def\etc{\hbox{\it etc}}\def\eg{\hbox{\it e.g.\ }}

\def\ie{\hbox{\it i.e.\ }}

\def\fiat{{}}
\def\\{\hfill\break} \def\={{ \; \equiv \; }}

\def\annota#1{\footnote{${{}^{\bf#1}}$}}
\ifnum\aux=1\BOZZA\else\relax\fi
\ifnum\tipoformule=1\let\Eq=\eqno\def\eq{}\let\Eqa=\eqno\def\eqa{}
\def\equ{{}}\fi
\def\defi{\,{\buildrel def \over =}\,}

\def\1{\ifnum\mgnf=0\pagina\else\relax\fi}
\def\W#1{#1_{\kern-3pt\lower6.6truept\hbox to 1.1truemm
{$\widetilde{}$\hfill}}\kern0pt}

\def\nn{{\V n}}\def\NN{{\cal N}}\def\LL{{\cal L}}
\def\aa{{\V\a}}
\def\EE{{\cal E}}\def\xx{{\V x}}\def\pps{{\V \ps}}
\def\cfr{{\it c.f.r.\ }}
\def\kk{{\V k}}\def\uu{{\V u}}
\def\VV#1{{\underline
#1}_{\kern-3pt$\lower7pt\hbox{$\widetilde{}$}}\kern3pt\,}

\def\FINE{
\*
\0{\it Internet:
Authors' preprints downloadable (latest version) at:

\centerline{\tt http://ipparco.roma1.infn.it}
\centerline{(link) \tt http://www.math.rutgers.edu/$\sim$giovanni}

\*
\sl e-mail: giovanni.gallavotti@roma1.infn.it
}}
\fiat
%**end of header

\fiat

\def\CH{chaotic hypothesis\ }

\centerline{\titolo Non equilibrium in statistical and fluid mechanics.}
\centerline{\bf Ensembles and their equivalence.
Entropy driven intermittency.}
\*\*
\centerline{\it Giovanni Gallavotti}
\*
\centerline{Fisica, Universit\`a di Roma 1}
\centerline{P.le Moro 2, 00185 Roma, Italia}
\*\*
\0{\bf Abstract:\it We present a review of the chaotic hypothesis and
discuss its applications to intermittency in statistical mechanics and
fluid mechanics proposing a quantitative definition. Entropy creation
rate is interpreted in terms of certain intermittency phenomena. An
attempt to a theory of the experiment of Ciliberto--Laroche on the
fluctuation law is presented.}
\*\*

\0{\bf\S1. Introduction.}
\numsec=1\numfor=1\*

A general theory of non equilibrium stationary phenomena extending
classical thermodynamics to stationary non equilibria is, perhaps
surprisingly, still a major open problem more than a century past
the work of Boltzmann (and Maxwell, Gibbs,...) which made the
breakthrough towards an understanding of properties of matter
based on microscopic Newton's equations and the atomic model.

In the last thirty years, or so, some progress appears to have been
achieved since the recognition that non equilibrium statistical
mechanics and stationary turbulence in fluids are closely related
problems and, in a sense, in spite of the apparently very different
nature of the equations describing them they are essentially the same.

The unifying principle, originally proposed for turbulent motions by
Ruelle, [Ru78], in the early 970's, has been extended to statistical
mechanics and eventually called the ``{\it chaotic hypothesis}'',
[GC95]:

\*
{\it Chaotic hypothesis: Asymptotic motions of a chaotic system, be it
a multi particle system of microscopic particles or a turbulent
macroscopic fluid, can be regarded as a transitive Anosov system for
the purposes of computing time averages in stationary states.}
\*

It may be useful to make a few comments on how this is supposed to be
interpreted. The conclusions that we draw here from the \CH are
summarized in \S13 which might be consulted at this point. For a
review on the subject seen from a different perspective see [Ru99a]
\*

{\bf \S2. Meaning of the chaotic hypothesis.}
\numsec=2\numfor=1
\*

Anosov systems are well understood dynamical systems: they play a
paradigmatic role with respect to chaotic systems parallel to the one
harmonic oscillators play with respect to orderly motions. They are so
simple, and yet very chaotic, that their properties are likely to be
the ones everybody develops in thinking about chaos, even without
having any familiarity with Anosov systems which certainly are not
(yet) part of the background of most contemporary
physicists.\annota{1}{\nota Informally a map $x\to Sx$ is a Anosov map
if at every point $x$ of the bounded phase space $M$ one can set up a
local system of coordinates with origin at $x$, continuously dependent
on $x$ and covariant under the action of $S$ and such that in this
comoving system of coordinates the point $x$ appears as a hyperbolic
fixed point for $S$. The corresponding continuous time motion, when
the evolution is $x\to S_t x, t\in R$, requires that the local system
of coordinates contains the phase space velocity $\dot x$ as one of
the coordinate axes and that the motion transversal to it sees $x$ as
a hyperbolic fixed point: note that a motion in continuous time cannot
possibly be hyperbolic in all directions and it has to be neutral in
the direction of $\dot x$ because the velocity has to be bounded if
$M$ is bounded, while hyperbolicity would imply exponential growth as
either $t\to+\io$ or $t\to-\io$. Furthermore there should be no
equilibrium points and the periodic points should be dense in phase
space. When the system has one or more (the so called ``{\it
hysteresis phenomenon''}) attracting sets which do not occupy the
whole phase space the \CH can be interpreted as saying that each
attracting set is a smooth surface on which the time evolution flow
(or map) acts as an Anosov flow (map).}

In general an Anosov system has asymptotic motions which approach one
out of finitely many invariant closed sets $C_1,\ldots,C_q$ each of
which contains a dense orbit,:one says that the systems $(C_j,S_t)$
are ``transitive''. One of them, at least, must be an attractive set.

To say that ``the asymptotic motions form a transitive Anosov
system'' means that
\*
\0(1) each of the sets $C_j$ which is attractive is a
smooth surface in phase space and
\\
(2) only one of them is attractive:
\*

The last ``transitivity'' assumption is meant to exclude the trivial
case in which there are more than one attractive sets and the system
{\it de facto} consists of several independent systems.

The smoothness of $C_j$ is a {\it strong assumption} that means that one
does not regard possible lack of smoothness, \ie fractality, as a
really relevant property in systems with large number of degrees of
freedom. In any event one could consider (if necessary) replacing
``Anosov systems'' with some slightly weaker property like ``axiom A''
systems which could permit more general asymptotic motions. Here we
adhere strictly to the chaotic hypothesis in the stated original form,
[GC95].
\*

{\bf\S3. Basic implications of the chaotic hypothesis and relation
with the ergodic hypothesis.}
\numsec=3\numfor=1\*

The chaotic hypothesis boldly extends to non equilibrium the {\it
ergodic hypothesis}: applied to equilibrium systems, \ie to systems
described by Hamiltonian equations, it implies the latter,
[Ga98]. This means that if a Hamiltonian system at a given energy is
assumed to verify the chaotic hypothesis, \ie to be a transitive
Anosov system, then for all observables $F$ (\ie for all smooth
functions $F$ defined on phase space)
$$T^{-1}\ig_0^T F(S_t x)\,dt\tende{T\to\io} \ig_M
F(y)\,\m_L(dy)\Eq(3.1)$$
where $\m_L$ is the Liouville distribution on the constant energy
surface $M$, and \equ(3.1) holds for almost all points $x\in M$, \ie
for $x$ outside a set $\NN$ of zero Liouville volume on $M$.

Being very general one cannot expect that the chaotic hypothesis will
solve any special problem typical of non equilibrium physics, like
``proving'' the Fourier's law of heat conduction, the Ohm's law of
electric conduction or the K41 theory of homogeneous turbulence.

Nevertheless, like the ergodic hypothesis in equilibrium, the \CH
accomplishes the remarkable task of giving us the ``statistics'' of
motions.  If $M$ is the phase space, which we suppose a smooth bounded
surface, and $t\to S_t x$ is the motion starting at $x\in M$, the time
average:

$$T^{-1}\ig_0^T F(S_t x)\,dt\tende{T\to\io} \ig_M
F(y)\,\m_{SRB}(dy)\Eq(3.2)$$
of the observable $F$ exists for $x$ outside a set $\NN$ of {\it zero
phase space volume} and it is $x$--independent, thus defining the
probability distribution $\m_{SRB}$ via \equ(3.2).

Note, in fact, that the probability distribution $\m_{SRB}$ defined by
the l.h.s. of \equ(3.2) is uniquely determined (provided it exists):
it is usually called the ``{\it statistics of the motion}'' or the
``{\it SRB distribution}'' associated with the dynamics of the system.

To appreciate the above property (existence and uniqueness of the
statistics) the following considerations seem appropriate.

An essential feature, and the main novelty, with respect to
equilibrium systems is that {\it non conservative forces} may act on the
system: this is in fact the very definition of ``non equilibrium
system''.

Since non conservative forces perform work it is necessary that on
the system act also other forces that take energy out of it,
at least if we wish that the system reaches a stationary state,
showing a well defined statistics.

As a consequence {\it any model} of the system must contain, besides
non conservative forces which keep it out of equilibrium by
establishing ``{\it flows}'' on it (like a heat flow, a matter flow,
...), also dissipative forces preventing the energy to increase
indefinitely and forcing the motion to visit only a finite region of
phase space.

The dissipation forces, also called ``thermostatting forces'', will in
general be such that the volume in phase space is {\it no longer}
invariant under time evolution. Mathematically this means that the
divergence $-\s(x)$ of the equations of motion will be not zero and
its time average $\ig_M \s(y)\m_{SRB}(dy)\defi\s_+$ will be positive
or zero as it cannot be negative (``because phase space is supposed
bounded'': see [Ru96]).

One calls a system ``{\it dissipative}'' if $\s_+>0$ and we expect
this to be the case as soon as there are non conservative forces
acting on it.

We see that if a system is dissipative then its statistics $\m_{SRB}$
{\it must} be concentrated on a set of {\it zero volume} in $M$: this
means that $\m_{SRB}$ cannot be very simple, and in fact it is
somewhat hard to imagine it.
\*

If the acting forces depend on a parameter $E$, ``strength of the non
conservative forces'', and for $E=0$ the system is Hamiltonian we have
a rather unexpected situation. At $E=0$ the \CH and the weaker ergodic
hypothesis imply that the statistics $\m_{SRB}$ is equal to the
Liouville distribution $\m_L$; but if $E\ne 0$, no matter how small,
it{\it will not} be possible to express $\m_{SRB}$ via some density
$\r_E(y)$ in the form $\m_{SRB}(dy)=\r_E(y)\m_L(dy)$, because
$\m_{SRB}$ attributes probability $1$ to a set $\NN$ with zero volume
in phase space (\ie $\m_L(\NN)=0$).  Nevertheless {\it natura non
facit saltus} (no discontinuities appear in natural phenomena) so that
sets that have probability $1$ with respect to $\m_{SRB}$ may be all
still dense in phase space, at least for $E$ small. In fact this is a
``{\it structural stability}'' property for systems which verify the
\CH (see [Ga96c])
\*

The above observations show one of the main difficulties of non
equilibrium physics: the unknown $\m_{SRB}$ is intrinsically more
complex than a function $\r_E(y)$ and we cannot hope to proceed in the
familiar way we might have perhaps expected from previous experiences:
namely to just set up some differential equations for the unknown
$\r_E(y)$.

Hence it is important that the \CH not only guarantees us the
existence of the statistics $\m_{SRB}$ but {\it also} that it does so
in a ``constructive way'' {\it giving at the same time formal
expressions for the distribution} $\m_{SRB}$ which should possibly play
the same role as the familiar formal expressions used in equilibrium
statistical mechanics in writing expectations of observables with
respect to the microcanonical distribution $\m_L$.
\*

For completeness we write a popular expression for $\m_{SRB}$. If $\g$
is a periodic orbit in phase space, $x_\g$ a point on $\g$, $T(\g)$
the period of $\g$ then

$$\ig F(y)\m_{SRB}(dy)=\lim_{T\to\io}
\fra{\sum_{\g:\,T(\g)\le T}
 e^{-\ig_0^{T(\g)}\s(S_t x_\g) \,dt}\,\ig_0^{T(\g)}F(S_t x_\g)\,dt}
{\sum_{\g:\,T(\g)\le T} e^{-\ig_0^{T(\g)}\s(S_t x_\g) \,dt}\,T(\g)}\Eq(3.3)
$$
This is simple in the sense that it does not require, to be formulated,
an even slight understanding of any of the properties of Anosov or
hyperbolic dynamical systems. But in many respects {\it it is not} a
natural formula: as one can grasp from the fact that it is far from
clear that in the equilibrium cases \equ(3.3) is an alternative
definition of the microcanonical ensemble (\ie of the Liouville
distribution $\m_L$), in spite of the fact that in this case $\s\=0$ and
\equ(3.3) becomes slightly simpler.

To prove \equ(3.3) one first derives alternative and much more useful
expressions for $\m_{SRB}$ which, however, require a longer discussion
to be formulated, see [Ga99a], [Ga86c]: the original work is due to
Sinai and in cases more general than Anosov systems, to Ruelle and Bowen.
\*
{\bf\S4. What can one expect from the chaotic hypothesis?}
\numsec=4\numfor=1\*

In equilibrium statistical mechanics we know the statistics of the
motions, if the ergodic hypothesis is taken for granted. However this
hardly solves the problems of equilibrium physics simply because
evaluating the averages is a difficult task which is also model
dependent. Nevertheless there are a few general consequences that can
be drawn from the ergodic hypothesis: the simplest (and first) is
embodied in the ``{\it heat theorem}'' of Boltzmann.

Imagine a system of $N$ particles in a box of volume $V$ subject to
pair interactions and to external forces with potential energy $W_V$,
due to the walls and providing the confinement of the particles to the
box. Define

$$\eqalign{
T=& \hbox{\rm average kinetic energy}\cr
U=& \hbox{\rm total energy}\cr
p=&\hbox{\rm  average of \ } \dpr_V W_V\cr}\Eq(4.1)$$
where the averages are taken with respect to the Liouville distribution
on the surface of energy $U$.

Imagine varying the parameters on which the system depends (\eg the
energy $U$ and the volume $V$) so that $dU,\, dV$ are the
corresponding variations of $U,V$, then

$$(dU+p\,dV)/T= \ {\rm exact}\Eq(4.2)$$
expresses the heat theorem of Boltzmann.

It is a consequence of the ergodicity assumption, but it is {\it
not} equivalent to it as it only involves a relation between a few
averages ($U,p,V,T$), see [Bo66], [Bo84], [Ga99a]. Not only it gives us
a relation which is a very familiar property of macroscopic systems,
but it also suggests us that even if the ergodic hypothesis is not
strictly valid some of its consequences might, still, be regarded
as correct.

The proposal is to regard the \CH in the same way: it is possible to
imagine that mathematically speaking the hypothesis is not strictly
valid and that, nevertheless, it yields results which are physically
correct for the few macroscopic observables in which one is really
interested in.

The ergodic hypothesis implies the heat theorem as a general
(``somewhat trivial'') {\it mechanical identity} valid for systems of
$N$ particles with $N=1,2,\ldots, 10^{23},\ldots$. For small $N$ it
might perhaps be regarded as a curiosity: such it must have been
considered by most readers of the key paper [Bo84] who were possibly
misled by several examples with $N=1$ given by Boltzmann in this and
other previous papers. Like the example of the system consisting of
one ``averaged'' Saturn ring, \ie one homogeneous ring of mass
rotating around Saturn with energy $U$, kinetic energy $T$ and
``volume'' $V$ (improbably identified with the strength of the
gravitational attraction!). But for $N=10^{23}$ it is no longer a
curiosity and it is a fundamental law of thermodynamics in
equilibrium: which, therefore, can be regarded on the same footing of
a symmetry being a direct consequence of the structure of the
equations of motion, [Ga99a] appendices to Ch.1 and Ch.9. It reflects in
macroscopic terms a simple microscopic assumption (\ie Newton's
equations for atomic motions, in this case).

No new consequences of even remotely comparable importance are known
to follow from the \CH besides the fact that it implies the validity
of the ergodic hypothesis itself (hence of all its consequences, first
of them classical equilibrium statistical mechanics).

Nevertheless the \CH {\it does} have some rather general
consequences. We mention here the {\it fluctuation theorem}. Let
$\s(x)$ be the phase space contraction rate and $\s_+$ be its SRB
average (\ie $\s_+=\ig \s(x)\,\m_{SRB}(dx)$), let $\t>0$ and define
$$p(x)=\t^{-1}
\ig_{-\t/2}^{\t/2}\fra{\s(S_tx)}{\s_+}\,dx\Eq(4.3)$$
and study the fluctuations of the observable $p(x)$ in the stationary
state $\m_{SRB}$. We write $\p_\t(p)\,dp$ the probability
that, in the distribution $\m_{SRB}$, the quantity $p(x)$ has actually
value between $p$ and $p+dp$ as

$$\p_\t(p) \,dp\,=\,{\rm const\ } e^{\z_\t(p)\,\t}dp\Eq(4.4)$$
Them $\lim_{\t\to\io} \z_\t(p)=\z(p)$ exists and is convex in $p$; and
\*

\0{\bf Theorem: \it (fluctuation theorem) Assume the \CH and suppose that
the dynamics is reversible, \ie that there is an isometry $I$ of phase
space such that

$$I S_t=S_{-t} I, \qquad I^2=1\Eq(4.5)$$
and that the attracting set is the full phase space.%
\annota{2}{\nota
It is perhaps important to stress that we distinguish between {\it
attracting set} and {\it attractor}: the first is a closed set such
that the motions that start close enough to it approach it ever
closer; an attractor is a subset of an attracting set that
\\
(1) has probability $1$ with respect to the statistics $\m$ of the
motions that are attracted by the attracting set (a notion which makes
sense when such statistics exists, but for a zero volume set of
initial data, and is unique) and that
\\
(2) has the smallest Hausdorff dimension among such probability $1$
sets. Hence density of an attracting set in phase space does not mean
that the corresponding attractor has dimension equal to that of the
phase space: it could be substantially lower, see [GC95].} Then

$$\z(-p)=\z(p)-\s_+\,p,\qquad {\rm for\ all\ } p\Eq(4.6)$$
where $\s_+=\m_{SRB}(\s)$.}
\*

It should be pointed out that the above relation was first discovered
in an experiment, see [ECM93], where also some theoretical ideas
were presented, correctly linking the result to the SRB distributions
theory and to time reversal symmetry. Although such hints were not
followed by what can be considered a proof, [CG99], still the
discovery has plaid a major role and greatly stimulated further
research.

The interest of \equ(4.6) is that, in general, it is a relation
without free parameters. The above theorem, proved in [GC95] for
discrete evolutions (maps) and in [Ge97] for continuous time systems
(flows), is one among the few general consequences of the chaotic
hypothesis, see [Ga96a], [Ga96b], [Ga99b] for others.
\*

{\bf\S5. Non equilibrium ensembles. Thermodynamic limits. Equivalence.}
\numsec=5\numfor=1
\*

The \CH gives us, unambiguously, the probability distribution
$\m_{SRB}$ which has to be employed to compute averages of observables
in stationary states.

For each value of the parameters on which the system depends we have,
therefore a well defined probability distribution $\m_{SRB}$. Calling
$\aa=(\a_1,\ldots,\a_p)$ the parameters and $\m_\aa$ the corresponding
SRB distribution we consider the collection
$\EE $ of probability distributions $\m_\aa$ obtained by letting the
parameters $\aa$ vary. We call such a collection an ``{\it
ensemble}''.

For instance $\aa$ could be the average energy $U$ of the system, the
average kinetic energy $T$, the volume $V$, the intensity $E$ of the
acting non conservative forces, \etc

Non equilibrium thermodynamics can be defined as the set of relations
that the variations of the parameters $\aa$ and of other average
quantities are constrained to obey as some of them are varied.  In
equilibrium the heat theorem is an example of such relations.  In
reversible non equilibria the fluctuation theorem \equ(4.6) is an
example.

In non equilibrium systems the equations of motion play a much more
prominent role than in equilibrium: in fact one of the main properties
of equilibrium statistical mechanics is that dynamics enters only
marginally in the definition of the statistical distributions of the
equilibrium states.

The necessity of a reversibility assumption in the fluctuation theorem
already hints at the usefulness of considering the equations of motion
themselves as ``parameters'' for the ensembles describing non
equilibrium stationary states: we are used to irreversible equations in
describing non equilibrium phenomena (like the heat equation, the Navier
Stokes equation, {\it etc}) and unless we are able to connect our
experiments with reversible dynamical models we shall be unable to make
use of the fluctuation theorem.

Furthermore it is quite clear that once a system is not in equilibrium
and thermostatting forces act on it, the exact nature of such forces
might be irrelevant within large equivalence classes: \ie it might be
irrelevant which particular ``cooling device'' we use to take heat out
of the system. Hence one would like to have a frame into which to set
up a more precise analysis of such arbitrariness. Therefore we shall
set
\*

\0{\bf Definition 1: \it A stationary ensemble $\EE$ for a system of
particles or for a fluid is the collection of $SRB$ distributions,
for given equations of motion, obtained by varying the parameters
entering into the equations.}
\*

It can happen that for the {\it same system} one can imagine {\it
different models}. In this case we would like that the models give the
same results, \ie the same averages to the same observables, at least
in some relevant limit. Like in the limit of infinite size in which
the number $N$, the volume $V$ and the energy $U$ tend to infinity but
$N/V$ and $U/V$ stay constant. Or in the limit in which the Reynolds
number $R$ tends to infinity in the case of fluids.

\def\LL{{\cal L}}
This gives the possibility of giving a precise meaning to the
equivalence of different thermostatting mechanisms. We shall declare

\*
\0{\bf Definition 2, \it  (equivalence of ensembles):
Two thermostatting mechanisms are equivalent ``in the thermodynamic
limit'' if one can establish a one to one correspondence between the
elements of the ensembles $\EE$ and $\EE'$ of SRB distributions
associated with the two models in such a way that the same observables,
in a certain class $\LL$ of observables, have the same averages in
corresponding distributions, at least when some of the parameters of
the system are sent to suitable limiting values to which we assign the
generic name of ``thermodynamic limit''.}
\*

In the following sections we illustrate possible applications of this
concept.
\*

{\bf\S6. Drude--Lorentz' electric conduction models.}
\numsec=6\numfor=1\*

Understanding of electric conduction is in a very unsatisfactory
state. It is usually based on linear response theory and very seldom a
fundamental approach is attempted. Of course this is so for a good
reason, because a fundamental approach would require imposing an
electric field $E$ on the system and, at the same time, a thermostatting
force to keep the system from blowing up and to let it approach a
steady state with a current $J_E$ flowing in it, and then taking the
ratio $J_E/E$ (with or without taking also the limit as $E\to0$).

However, as repeatedly mentioned, it is an open problem to study
steady states out of equilibrium. Hence most theories have recourse to
linear response  where the problem of studying stationary non
equilibria does not even arise.

The reason why this is unsatisfactory is that as long as we are {\it in
principle} unable to study stationary non equilibria we are also {\it
in principle} unable to estimate the size of the approximation and
errors of linear response.

In spite of many attempts the old theory of Drude, see [Be59], [Se87],
seems to be among the few conduction theories which try to establish a
conductivity theory based on the study of electric current at
non zero fields.

We imagine a set of obstacles distributed randomly or periodically and
among them conduction electrons move, roughly with density of one per
obstacle.

The (screened) interactions between the electrons are, at a first
approximation, ignored. The collisions between electrons and obstacles
(``nuclei'') will take place in the average after the electrons have
traveled a distance $\l=(\r a^2)^{-1}$ if $\r$ is the nuclei density
and $a$ is their radius.

Between collisions the electrons, with electric charge $e$, accelerate
in the direction of an imposed field $\V E$ incrementing, in that
direction, velocity by

$$\d v=\fra{e E \l}{m v}=\fra{e E (\r a^2)^{-1}}{m
\sqrt{k_BT/m}}\Eq(6.1)$$
where $k_B$ is Boltzmann's constant. At collision they are
``thermalized'': an event that is modeled by giving them a new velocity
of size $v=\sqrt{k_B T/m}$ and a random direction.

The latter is the ``thermostatting mechanism'' which is a, somewhat
rough, description of the energy transfer from electrons to
lattice which physically corresponds to electrons losing energy in
favor of lattice phonons, which {\it in turn} are kept at constant
temperature by some other thermostatting mechanism which prevents the
wire melting. All things considered the total current that flows will
be

$$J_E= \fra{e^2}{\r a^2 \sqrt{m k_B T}}\,E\defi\ch\, E\Eq(6.2)$$
obtaining Ohm's law.

To the same conclusion we arrive by a different thermostat model. We
imagine that the electrons move exchanging energy with lattice phonons
but keeping their {\it total energy} constant and equal to $N\,k_B\, T$:
\ie $2^{-1}\sum_{j=1}^N m \dot{\V x}_j^2=3 N k_B T/2$, where $k_B$ is
Boltzmann's constant. There are several forces that can achieve this result

we select the ``Gaussian minimal constraint'' force.\annota{3}{\nota
Not because it plays any fundamental role but because it has been
studied by many authors and because it represents a mechanism very
close to that proposed by Drude.We recall, for copleteness, that the
{\it effort} of a constraint reaction on a motion on which the active
force is $\V f$ (with $3N$ components) and $\V a$ is the acceleration
of the particles (with $3N$ components) and $m$ is the mass is $\EE(\V
a)=\,(\V f-m\,\V a)^2/m$; then Gauss' principle is that the effort is
minimal if $\V a$ is given the actual value of the acceleration, at
fixed space positions and velocities.}  This is the force that is
required to keep $\sum m
\dot{\V x}^2_j$ strictly constant and that is determined by ``Gauss
least effort'' principle, see [Ga99a], ch. 9, appendix 4, for instance:
as is well known this is, on the $i$--th particle, a force

$$-\a\,\dot{\xx}_i\defi -\fra{e \V E\cdot\sum_j\dot{\xx}_j}{\sum_j
\dot\xx_j^2}
\dot \xx_i\= -\,\fra{m \V E \cdot N\,\V J}{3 N k_B T}
\,\dot\xx_i\Eq(6.3)$$
If there are $N$ particles and $N$ is large it follows that $\V
J=N^{-1}e\sum_j\dot\xx_j$ is essentially constant, see [Ru99b], and
each particle evolves, almost independently of the others, according to
an equation:

$$m \ddot\xx_i=e\V E-\n \dot\xx_i\Eq(6.4)$$
between collisions, with a suitably fixed constant $\n$. If we imagine
that the velocity of the particles between collisions changes only by a
small quantity compared to the average velocity the ``friction term''
which in the average will be of order $E^2$ will be negligible {\it
except} for the fact that its ``only'' effect will be of insuring that
the {\it total} kinetic energy stays constant and the speeds of the
particles are constantly renormalized. In other words this is the same
as having continuously collisions between electrons and phonons even
when there is no collision between electrons and obstacles. Hence the
resulting current is the same (if $N$ is large) as in \equ(6.2).
\*

{\bf\S7. Ensemble equivalence: the example of electric conduction
theories.}
\numsec=7\numfor=1\*

We have derived three models for the conduction problem,
namely
\*

(1) the classical model of Drude, [Se87], in which at {\it every
    collision} the electron velocity is reset to the average velocity
    at the given temperature, with a random direction, \cfr \equ(6.1)
    and \equ(6.2).

(2) the Gaussian model in which the total kinetic energy is kept
    constant by a thermostat force

$$ m \ddot \xx_i=
\V E
-\,\fra{m \V E \cdot\,\V J}{3 k_B T}
\,\dot\xx_i+ ``{\rm collisional\ forces}''\Eq(7.1)$$
where $3 N k_B T$ is the total kinetic energy (a constant of motion in
this model). The model has been widely studied and it was introduced
by Hoover and Evans (see for instance [HHP87] and [EM90]).

(3) a ``friction model'' in which particles independently
    experience a constant friction

$$ m \ddot \xx_i=
\V E-\,\n \,\dot\xx_i+ ``{\rm collisional\ forces}''\Eq(7.2)$$
where $\n$ is a constant tuned so that the {\it average kinetic
energy} is $e N k_B T/2$. This model was considered in the perspective
of the conjectures of ensemble equivalence in [Ga95],
[Ga96b].
\*

The first model is a ``stochastic model'' while the second and third
are deterministic: the third is ``irreversible'' while the second is
reversible because the involution $I(\xx_i,\V v_i)=(\xx_i,-\V v_i)$
anticommutes with the time evolution flow $S_t$ defined by the
equation \equ(7.1): $I S_t=S_{-t}I$ (as the ``friction term'' is {\it
odd} under $I$).

Let $\m_{\d,T}$ be the SRB distribution for \equ(7.1) for the
stationary state that is reached starting from initial data with
energy $3N k_B T/2$. The collection of the distributions $\m_{\d,T}$
as the kinetic energy $T$ and the
density $\d=N/V$ vary, define a ``statistical ensemble'' $\EE$ of
stationary distributions associated with the equation \equ(7.1).

Likewise we call $\tilde \m_{\d,\n}$ the class of SRB distributions
associated with \equ(7.2) which forms an ``ensemble'' $\tilde \EE$.

We establish a correspondence between distributions of the ensembles
$\EE$ and $\tilde \EE$: we say that $\m_{\d,T}$ and $\tilde
\m_{\d',\n}$ are ``corresponding elements'' if

$$\d=\d',\qquad T=\ig \fra12(\sum_j m \dot\xx^2_j)\,\tilde
\m_{\d,\n}(d\xx\,d\dot\xx)
\Eq(7.3)$$
Then the following conjecture was proposed in [Ga96b].
\*

{\bf Conjecture 1: \it (equivalence conjecture) Let $F$ be a ``{\sl local
observable}'', \ie an observable depending solely on the microscopic
state of the electrons whose positions is inside a  fixed box $V_0$.
Then, if $\LL$ denotes the local smooth observables

$$\lim_{N\to\io, N/V=\d} \tilde \m_{\d,\n}(F)=
\lim_{N\to\io, N/V=\d} \m_{\d,T}(F) \qquad F\in \LL\Eq(7.4)$$
if $T$ and $\n$ are related by \equ(7.3).}
\*

This conjecture has been discussed in [Ga95], sec. 5, and [Ga96a], see
sec. 2 and 5: and in [Ru99b] arguments in favor of it have been
developed.

Clearly the conjecture is very similar to the equivalence in
equilibrium between canonical and microcanonical ensembles: here the
friction $\n$ plays the role of the canonical inverse temperature and
the kinetic energy that of the microcanonical energy.

It is remarkable that the above equivalence suggests equivalence
between a ``reversible statistical ensemble'', \ie the collection
$\EE$ of the SRB distributions associated with \equ(7.1)  and a
``irreversible statistical ensemble'', \ie the collection $\tilde \EE$
of SRB distributions associated with \equ(7.2).
\*

Furthermore it is natural to consider also the collection $\EE'$ of
stationary distributions for the original stochastic model (1) of Drude,
whose elements $\m'_{\n,T}$ can be parameterized by the quantities $T$,
temperature (such that $\fra12\sum_j m
\dot\xx_j^2=\fra32 N k_B T$, and $N/V=\d$). This is an ensemble $\EE'$
whose elements can be put into one to one correspondence with the
elements of, say, the ensemble $\EE$ associated with model (2), \ie
with \equ(7.1): an element $\m'_{\n,T}\in \EE'$ corresponds to
$\m_{\d,\n}\in \EE$ if $T$ verifies \equ(7.3). Then
\*

{\bf Conjecture 2: \it If $\m_{\d,T}\in \EE$  and $\m'_{\d,\n}\in \EE'$
are corresponding elements (\ie \equ(7.3) holds) then
$$\lim_{N\to\io, N/V=\d} \m_{\d,T}(F)=
\lim_{N\to\io, N/V=\d} \m'_{\d,T}(F) \qquad F\in \LL\Eq(7.5)$$
for all local observables $F\in \LL$.}
\*

Hence we see that there can be statistical equivalence between a
viscous irreversible dissipation model and either a stochastic
dissipation model or a reversible dissipation model, at least as far
as the averages of special observables are concerned.

The argument in [Ru99b] in favor of conjecture 1 is that the
coefficient $\a$ in \equ(6.3) is essentially the average $J$ of the
current over the {\it whole} box containing the system of particles, $J=
N^{-1}\,e\,\sum_j \dot\xx_i$: hence $J$ should be constant with probability
$1$, at least if the stationary SRB distributions can be reasonably
supposed to have some property of ergodicity with respect to {\it space
translations}.
\*

{\bf\S8. Entropy driven intermittency in reversible dissipation.}
\numsec=8\numfor=1\*

A further argument for the equivalence conjectures in the above
electric conduction models can be related to the fluctuation theorem:
the quantity $\a(x)$ is also proportional to the phase space
contraction rate $\s(x)=(3N-1)\a(x)$. Therefore, denoting in general
with a subscript $+$ the SRB average (or the time average) of an
observable, the probability that $\s(x)$ deviates from its average
$\s_+=(3N-1)\,\a_+$ can be studied as follows.

If the number $N$ of particles is large the time scale $\t_0$ over
which $\s(S_t x)$ evolves will be large compared to the microscopic
evolution rates, because $\s_t(x)$ is the {\it sum} of the $\sim 6N$
rates of expansion and contraction of the $\sim 6N$ phase space
directions out of $x$ (sometimes called the ``{\it local Lyapunov
exponents}'').\annota{4}{\nota The exact number of exponents depends
on how many constants of motion the system has: for instance in the
case of the conduction model (1) in \S6 above the number of exponents
is $6N-1$ because the kinetic energy is conserved and the system has
no other (obvious) first integrals. Furthermore one of such exponents is
$0$ since every dynamical system in continuous time has one zero
exponent (corresponding to the direction $\dot x$ of the flow).}
\*

Consider a large number $m$ of time intervals $I_1,I_2,\ldots,I_m$ of
size $\t_0$ and let $\s_j$ be the (average) value of $\s(S_t x)$ for
$t\in I_j$. Then the fraction of the $j$'s such that $\s_j-\s_+\simeq
\s_+ p$ will be proportional to

$$\p_{\t_0}(p)\simeq e^{\t_0\z(p)} \Eq(8.1)$$
and $\z(p)<\z(1)$  if $p\ne1$. Since we can expect that$\z(p)$ is
proportional to $N$ we
see that the fraction of time intervals $I_j$ in which $\s_j\ne\s_+$
will be exponentially small with $N$.

For instance the fraction of time intervals in which $\s_j\simeq -\s_+$
will be, by the fluctuation theorem

$$e^{-(3N-1)\a_+ \t_0}\Eq(8.2)$$
In order that the above argument holds it is essential that $N$ is
large to the point that we can think that the time scale $\t_0$ over
which $\s(S_t x)$ varies is much larger than the microscopic scales:
so that we can regard $\t_0$ large enough for the fluctuation theorem
to apply. In this respect this is not really different from the
previously quoted argument in [Ru99b]. However the change of
perspective gives further information.
\*

In fact we get the following picture: $N$ is large and for most of the
time the (stationary) evolution uneventfully proceeds as if $\s(S_t x)\=
\s_+$ (thus justifying conjecture 1). Very rarely, however, it proceeds
as if $\s(S_t x)\ne\s_+$, for instance with $\s(S_tx)=-\s_+$: such
``{\it bursts of anomalous behavior}'' occur very rarely. But when
they occur ``everything else goes the wrong way'' because, as
discussed in detail in [Ga99c], while the phase space contraction is
opposite to what it ``should be'' (in the average) then it also happens that
{\it all observables evolve following paths that are the time reversal
of the expected paths}, This is the content, see [Ga99c], of the
following theorem which is quite close (particularly if one examines
its derivation) to the Machlup--Onsager theory of fluctuation patterns
(note that, however, it does not require closeness to equilibrium)
\*

\0{\bf Theorem\it (conditional reversibility theorem): If $F$ is an
observable with even (or odd), for simplicity, time reversal parity and
if $\t$ is large then the evolution or ``fluctuation pattern'' $\f(t)$
and its time reversal $I\f(t)\=\f(-t)$, $t\in [-\t_0/2,\t_0/2]$, will be
followed with equal likelihood if the first is conditioned to an entropy
creation rate $p$ and the second to the opposite $-p$.}
\*

In other words systems with reversible dynamics can be equivalent to
systems with irreversible dynamics but they show ``{\it intermittent
behavior}'' with intermittency lapses that become extremely rare very
quickly as $N\to\io$. Sometimes they can be really dramatic, as in the
cases in which $\s=-\s_+$: alas they are unobservable just for this
reason and one can wonder (see \S9 below) whether this is really of
any interest.
\*

{\bf\S9. Local fluctuations and observable intermittency.}
\numsec=9\numfor=1\*

As a final comment upon the analysis of the equivalence of ensembles
attempted above we consider a very large system with volume $V$ and a
small subsystem of volume $V_0$ which is large but not yet really
macroscopic so that the number of particles in $V_0$ is not too large,
a nobler way to express the same notion is to say that
we consider a ``mesocopic'' subsystem of our macroscopic system.

Here it is quite important to specify the system because we want to
make use of aspects of the equivalence conjectures that are model
dependent.  Therefore we consider the conduction models (2) or (3) of
\S5: these are models in which dissipation occurs ``homogeneously''
throughout the system. In this case we can imagine to look at the part
of the system in the box $V_0$: if $j_1,\ldots,j_{N_0}$ are the
particles which at a certain instant are inside $V_0$ and
$\dot\xx_j=\V f_j(\xx)$ are the equations of motion we can define

$$\s_{V_0}(x)=\sum_{i=1}^{N_0} \dpr_{x_{j_k}} f_{j_k}(x)\Eq(9.1)$$
which is (by definition) the part of phase space contraction due to
the particles in $V_0$.

Since the part of the system inside the microscopically large but
macroscopically small $V_0$ can be regarded as a new dynamical system
whose properties should not be different from the ones of the full
system enclosed in the full volume $V$ we may expect that the subsystem
inside $V_0$ is in a stationary state and the quantity $\s_{V_0}$ has
the same fluctuation properties as $\s_V$,
\ie

$$\eqalign{
(1)&\qquad \media{\s_{V_0}}_+= V_0\, \lis \s_+,
\qquad \media{\s_{V}}_+= V\, \lis \s_+\cr
(2)&\qquad \p_\t^{V_0}(p)= e^{\lis \z(p)\,\t\,V_0} ,
\qquad \qquad \p_\t^{V}(p)= e^{\lis \z(p)\,\t\,V} \cr}\Eq(9.2)$$
where $\lis\z,\lis\s_+$ are {\it the same for} $V,V_0$ and $p=\t^{-1}
\ig_{-\t/2}^{\t/2} \s_{V_0}(S_t x)/\media{\s_{V_0}}\,dt$ or
respectively $p=\t^{-1}
\ig_{-\t/2}^{\t/2} \s_{V}(S_t x)/\media{\s_{V}}\,dt$. Here $\s_{V_0}$
is naively defined as the contribution to $\s$ coming from the
particles in $V_0$.
\*

{\it In other words in large stationary systems with homogeneous
reversible dissipation phase space contractions fluctuate in an {\sl
extensive way}, \ie they are regulated by the {\sl same} deviation
function $\lis\z(p)$} (volume independent).
\*

This is very similar to the well known property of equilibrium density
fluctuations in a gas of density $\r$: if $V\supset V_0$ are a very
large volume $V$ in a yet larger container and $V_0$ is a small but
microscopically large (\ie mesoscopic) volume $V_0$ then the total
numbers of particles in $V$ and $V_0$ will be $N$ and $N_0$ and the
average numbers will be $\r V$ and $\r V_0$ respectively. Then setting

$$p=(N-\r\,V)/\r V,\qquad or, \qquad p=(N_0-\r\,V_0)/\r V_0\Eq(9.3)$$
the probability that the variable $p$ has a given value will be
proportional to

$$\p^V(p)=e^{\lis\z(p) V},\qquad \p^{V_0}(p)=e^{\lis\z(p)
V_0}\Eq(9.4)$$
{\it again with the same function} $\lis\z(p)$.
\*

This means that we {\it can observe $\lis \z(p)$ by performing
fluctuations experiments in small boxes}, ideally carved out of the
large container, where the density fluctuations are not too rare.
A ``local fluctuation law'' should hold more generally in cases of
models in which dissipation occurs homogeneously across the system,
like the above considered conduction models.

The intuitive picture for the above ``local fluctuation relation''
inspired (and was substantiated) a mathematical model in which a local
fluctuation relation can be proved as a theorem: it has een discussed
in [Ga99c], see also below.

Going back to the conduction model we see that the intermittency
phenomena discussed above can be actually observed by looking at the
fluctuations of the contribution to phase space contraction due to a
small subsystem.

And such ``{\it entropy driven}'' intermittency will be model
independent for models which are equivalent in the sense of the
previous sections provided the models used are equivalent and one of
them is reversible.

An extreme case is provided by models (1)\%(3), \S 7, for electric
conduction (conjectured to be equivalent, see \S7). In fact at first the
model (3), the viscous thermostat, might look uninteresting as,
obviously, in this case

$$\s^V(x)=\,3 N\n,\qquad \s^{V_0}(x)=\,3\,N_0\,\n\Eq(9.5)$$
and $\s^V/V$ has no fluctuations.

However the equivalence conjecture makes a statement about expectation
values of the {\it same observable}: hence we should consider the
quantity $\hat \s^{V_0}(x)=\V E\cdot \V J_{V_0}/ \sum_j \dot\xx_j^2$
and we should expect that its statistics with respect to an element of
the ensemble $\EE'$ is the same as that of the same quantity with
respect to the corresponding elements of the ensembles
$\EE,\tilde\EE$. Hence in particular the functions $\lis\z(p)$ which
control the large fluctuations of $\s^V(p)$ will verify

$$\lis \z(-p)=\lis\z(p)-p \media{\hat\s^{
V_0}}_+/V_0=\lis\z(p)-3\r\n\,p
=\lis \z(p)-\fra{ e E m\,J_+}{k_B T}\,p\Eq(9.6)$$
where the first equality expresses the validity of a fluctuation theorem
type of relation due to the fact that the small system, by the
equivalence conjecture, should behave as a closed system; the second
equality expresses a consequence of the equivalence conjecture between
models (2) and (3) while the third is obtained by expressing the
current via Drude's theory (again assuming the conjectures of
equivalence 1,2 of \S7).
\*

{\bf\S10. Fluids.}
\numsec=10\numfor=1\*

The \CH was originally formulated to understand developed turbulence,
[Ru78]: it is therefore interesting to revisit fluid motions theory.

The incompressible Navier Stokes equation for a velocity field $\V u$
in a periodic container $ V$ of side $L$ can be considered as an
equation for the evolution in time of its Fourier coefficients
$\uu_\kk$ where the ``{\it mode}'' $\kk$ has the form $2\p L^{-1}\V n$
with $\V n\ne\V0$ and $\nn$ an integer components
vector.\annota{5}{\nota The value $\V n=\V0$ is excluded because,
having periodic boundary conditions, it is not restrictive to suppose
that the space average of $\V u$ vanishes (galilean invariance). The
convention for the Fourier transform that we use is $\V u(\V
x)=\sum_\kk e^{i\,\kk\cdot\V x} \V u_\kk$.}  Furthermore
$\uu_\kk=\lis\uu_{-\kk}$ and $\kk\cdot\uu_\kk\=0$. If $p$ is the
pressure field and $\V f$ a simple forcing we shall fix the ideas by
considering $\V f(\xx)=f\,\V e \,\sin \kk_f\cdot\xx$ where $\kk_f$ is
some prefixed mode and $\V e$ is a unit vector orthogonal to $\kk_f$.

The Navier Stokes equation is then
$$\dot{\uu}+\W u\cdot\W\dpr\, \uu=-\Dpr p+\V f +\n\D\V u\Eq(10.1)$$
and it is convenient to use {\it dimensionless variables} $\V u_0,p_0,
\V\f_0,\V\x,\t$: so we define them as

$$\eqalign{
\uu(\xx,t)=& f L^2\n^{-1}\uu_0(L^{-1}\xx,L^{-2}\n t),\qquad
\V\x=L^{-1}\xx,\ \t=L^{-2}\n t\cr
p(\xx,t)=& f L p_0(L^{-1}\xx,L^{-2}\n t), \kern1.5cm R\defi f\,
L^3\,\n^{-2}\cr
\V f(\xx,t)=& f \V\f_0(L^{-1}\xx)\cr}\Eq(10.2)$$
with $\max |\V\f_0|=1$.  The result, {\it dropping the label $0$} and
calling again $\xx,t$ the new variables $\V\x,\t$, is that  the Navier
Stokes equations become an equation for a divergenceless field $\uu$
defined on $ V=[0,1]^3$, with periodic boundary conditions and equations
$$\dot{\uu}+R \,{\W u\cdot\W\dpr}\, \uu=-\Dpr p+\V \f
+\D\V u,\qquad \Dpr\cdot \V u=0\Eq(10.3)$$
with $\max |\V\f|=1$.

Equation \equ(10.3) is our model of fluid motion, where $R$ plays the
role of ``forcing intensity'' and the term $\D \uu$ represents the
``thermostatting force''.  As $R$ varies the stationary distributions
$\m_R$ which describe the SRB statistics of the motions \equ(10.3)
define a set $\EE$ of probability distributions which forms an
``ensemble''.

The mathematical theory of the Navier Stokes equations is far from being
understood: however phenomenology establishes quite clearly a few
key points.  The main property is that if \equ(10.3) is written as an
equation for the Fourier components of $\uu$ then one can assume that
$\uu_\kk\=\V0$ for $|\kk|> K(R)$, for some finite $K(R)$.

Therefore the equation \equ(10.3) should be thought of as a ``truncated
equation'' in momentum space by identifying it with the equation
obtained by projecting also $\W u\cdot\W\dpr\,\uu$ on the same function
space.
\*

Should one develop anxiety about the mathematical aspects of the
Navier Stokes equation one should therefore think that an equally good
model for a fluid is the mentioned truncation {\it provided $K(R)$} is
chose large enough.

The idea is that for $K(R)=R^\k$, with $\k$ larger than a suitable
$\k_0$ the results of the theory, \ie the statistical properties of
$\m_R$ become $\k$--independent for $R$ large.

The simplest evaluation of $\k_0$ gives $\k_0=9/4$ as a consequence of
the so called K41 theory of homogeneous turbulence, see [LL71].

If \equ(10.3) is a good model for a fluid when $L$ is large then it
provides us with an ``ensemble'' $\EE$ of SRB distributions (on the
space of the velocity fields components $\uu_\kk$ of dimension $\sim
8\p K(R)^3/3$).\annota{6}{\nota There are about $4\p K(R)^3/3$ vectors
with integer components inside a sphere of radius $K(R)$, thus the
number of complex Fourier components with mode label $|\kk|<K(R)$
would be $3$ times as much, but the divergenceless condition leaves
only $2$ complex components for $\uu_\kk$ along the two unit vectors
orthogonal to $\kk$ and the reality condition further divides by $2$
the number of ``free'' components.}
\*

We should expect, following the discussion of the statistical mechanics
cases, that there can be other ``ensembles'' $\tilde \EE$ which are
equivalent to $\EE$.

Here $R$ plays the role of the volume in non equilibrium statistical
mechanics, so that $R\to\io$ will play the role of the thermodynamic
limit, a limit in which the effective number of degrees of freedom,
$\sim4\p R^{3\k}/3$, becomes infinite. The role of the local
observables will be plaid by the (smooth) functions $F(\uu)$ of the
velocity fields $\uu$ which depend on $\uu$ only via its Fourier
components that have mode $\kk$ with $|\kk|<B$ for some $B$:
$F(\uu)=F(\{\uu_\kk\}_{|\kk|\le B})$.

We shall call $\LL$ the space of such observables: examples can be
obtained by setting $F(\uu)=|\ig e^{i\kk\cdot\xx} \uu(\xx)\,d\xx|^2$
or $F(\uu)=\ig
\pps(\xx)\cdot\uu(\xx)\,d\xx$ where the function has only a finte
number of harmonics,  $\pps(\xx)=\ig \sum_{|\kk|<B} e^{ i\kk\cdot\xx}
\uu(\xx)\,d\xx$, \etc.
\*

As in non equilibrium statistical mechanics we can expect that the
equations of motion themselves become part of the definition of the
ensembles. For instance one can imagine defining the ensemble $\tilde
\EE$ of the SRB distributions $\tilde \m_V$ for the equations

$$\dot\uu+ R\,\W u\cdot\W \dpr\uu=-\Dpr p+\pps+ \n(\uu)\D\uu\Eq(10.4)$$
called GNS equations in [Ga97a], or ``gaussian Navier Stokes'' equations,
where $\n(\uu)$ is so defined that

$$\X=\ig_V(\W\dpr\uu)^2 \,d\xx/(2\p)^3=\sum_\kk \kk^2\,|\uu_\kk|^2
\Eq(10.5)$$
is exactly constant and equal to $\X$. The equations \equ(10.4) are
interpreted as above with the same momentum cut off $K(R)=R^\k$.

An element $\tilde \m_\X$ of $\tilde \EE$ and one $\m_R$ of $\EE$, SRB
distributions for the two different dynamics \equ(10.3) and \equ(10.4),
``correspond to each other'' if

$$\X= \ig \m_R(d\V u)\,\big(\ig_V(\W\dpr\uu)^2
\,d\xx/(2\p)^3\big)\defi \X_R\Eq(10.6)$$
where $\m_R\in\EE$ is the SRB distribution at Reynolds number $R$ for
the previous viscous Navier Stokes equation, \equ(10.3), and we
naturally conjecture
\*

{\bf Conjecture 3\it (equivalence GNS--NS):
If $R\to\io$ then for all local observables $F\in \LL$ it is
$\m_R(F)=\tilde \m_{\X_R}(F)$ if \equ(10.6) holds.}
\*

It is easy to check that the GNS model ``viscosity'' $\n(\uu)$, having to
be such that the quantity $\X$ in \equ(10.5) is exactly constant must be

$$\n(\uu)=\fra{\ig_V\big(\V\f\cdot\D\uu-R\D\uu\cdot(\W
u\cdot\W\dpr\uu)\big)\,d\xx} {\ig_V(\D\uu)^2\,d\xx}\Eq(10.7)$$
and we see that while \equ(10.3) is an {\it irreversible} equation the
\equ(10.4) is {\it reversible}, with time reversal symmetry given by

$$I \uu(\xx,t)=- \uu(\xx,t)\Eq(10.8)$$
as one can check.

More generally one may wish to leave the ``Kolmogorov parameter'' $\k$
as a free parameter: in this case the SRB distributions will form an
ensemble whose elements can be parameterized by $R,\k$ and the
equivalence conjecture can be extended to this case yielding equivalence
between $\m_{R,\k}$ and $\tilde\m_{\X,\k}$. This is of interest,
particularly if one has numerical experiments in mind.
\*

If $\k>\k_0$ then the value of $\k$ {\it should be irrelevant}: but if
$\k<\k_0$ the phenomenology will be different from the one of the
Navier Stokes equation and equivalence might still hold but one cannot
expect either equation to have the properties that we expect for the
usual Navier Stokes equations (\ie in this situation one would have to
be careful in making statements based on common experience).

If we take $\k$ to be exactly equal to the value $\k_0=9/4$ (\ie if we
take the ultraviolet cut--off to be such that, according to the K41
theory, for larger values it is {\it needlessly large} and for lower
values it is {\it incorrectly low} and shows a phenomenology which
will depend on its actual value) then we may speculate that the
``attracting set'' is the full phase space (available compatibly with
the constraint $\X=\X_R$). Therefore the divergence of the equations of
motion, which is given by a rather involved expression in which only
the first term seems to dominate at large $R$, namely

$$
\eqalign{
&\s(\uu)=(\sum_{|\kk|< K(R)}\kk^2)\, \n(\uu)-\big(\ig_V
\D\V\f\cdot\D\uu\,d\xx\big)\,\big(\ig_V [(\D\uu)^2-\cr
&- R\, \D\uu\cdot(\D(\W u\cdot\W\dpr \uu))-R (\D \W
u)\cdot(\D\uu)\cdot (\W\dpr \uu) -R\D\uu\cdot (\D\W\dpr \uu)\W u+\cr
&+\n(\uu)\D\uu\cdot \D^2\uu ]d\xx\,\big)\ / \ig_V
(\D\uu)^2\,d\xx\cr}\Eq(10.9)$$
will verify the fluctuation theorem, \ie the rate function $\z(p)$ for
the average phase space contraction
$p=\t^{-1}\ig_{-\t/2}^{\t/2}\s(S_t\uu)dt/\t \s_+$ will be such that
$\z(-p)=\z(p)-p\s_+$.
\*

{\it If the \CH is valid together with the equivalence conjecture} the
validity of the fluctuation relation can be taken as a criterion for
determining $\k$: it would be the last $\k$ before which the
fluctuation relation between $\z(p)$ and $\z(-p)$ holds. {\it However
this conclusion can only be drawn if the attracting set in phase space
is the full ellipsoid $\X=\X_R$} at least for $K(R)=R^{\k_0}$.
\*

The latter property might not be realized: and in such case the
fluctuation theorem does not apply directly, although the equivalence
conjectures still hold. In fact one can try to extend the fluctuation
theorem to cover reversible cases in which the attracting set is
smaller than the full phase space left available by the
constraints. In such cases {\it under suitable geometric assumptions},
[BG97] and the earlier work [BGG97], one can derive a relation like

$$\z(-p)=\z(p)- p\s_+ \,\th, \qquad 0\le\th\le 1\Eq(10.10)$$
where $\th$ is a coefficient that can be related to the Lyapunov
spectrum of the system, \cfr [BG97], [Ga97a]. In fact numerical work
to check the theory proposed in [Ga97a] is currently being performed
(private communication by Rondoni and Segre)
with not too promising results which, optimistically, can be
attributed to the fact that the ultraviolet cut off is too small due
to numerical litmitations: clearly there is more work to do here. The
preliminary numerical results give, so far, the {\it somewhat
surprising linearity} in $p$ but with a slope that, although of the
correct order of magnitude, seems to have a value that does not match
the theory within the error bounds.
\*

Coming back to the Navier Stokes equation we mention that we may
imagine to write it as \equ(10.3) {\it but} with the different
constraint

$$U=\ig_V \uu^2 \,d\xx=\,{\rm const}\Eq(10.11)$$
rather than \equ(10.5).

This case has been considered in [RS99] and
the multiplier $\n(\uu)$ is in this case

$$\n(\uu)=\fra{\ig_V \V\f\cdot\uu\, d\xx}{\ig_V \uu^2\,d\xx},\qquad
\s(\uu)=(3\sum_{|\kk|<K(R)}\,|\kk|^2-1)\, \n(\uu)\Eq(10.12)$$
and we can (almost) repeat the above considerations and
equivalence conjectures. This constraint is a gaussian constraint that
$U$ is constant obtained by imposing its constancy on the Euler
evolution via Gauss' principle with a suitable definition of the notion
of ``constraint effort'' (this notion is not unique, see [Ga97a] for
another definition) and we do not discuss it here to avoid overlapping
with \S12 below.
\*

Thd intuitive motivation for the equivalence conjectures is that for
large $R$ the phase space contraction $\s(\uu)$ and the coefficient
$\n(\uu)\,$\annota{7}{\nota Which in the case \equ(10.9) are simply
proportional and in the case of \equ(10.4) they are related in a more
involved way, see
\equ(10.8),\equ(10.9), but which are still probably proportional to leading
order as $R\to\io$.} are ``global quantities'' and depend on the
global properties of the system (\eg $\s(\uu)$ is the sum of all the
local Lyapunov exponents of the system whose number is $O(K(R)^3)$):
they will ``therefore'' vary over time more slowly than any time scale
of the system and can be considered constant. 

The argument is not very convincing in the case of the equations with
the constraint \equ(10.11) because the $\s(\uu)$ in \equ(10.12) is
proportional to $\ig_V\V\f\cdot\uu\,d\xx$ which clearly depends {\it
only} on harmonics of $\uu$ with $\kk$ small, \ie it is a ``local
observable''. Note that this does not apply to the GNS equations with
the constrained vorticity $\X$, \equ(10.6) where the ``main''
contribution to $\s(\uu)$, see \equ(10.7), comes from the term
proportional to $R$ which contains all harmonics. Therefore the result
in [RS99] about the equivalence between the GNS equations, \equ(10.4)
with the constraint \equ(10.5), and the equations with constraint
\equ(10.11) is interesting and puzzling: it might be an artifact of
the smallness of the cut off that one has to impose in order to have
numerically feasible simulations.
\*

Finally $\s_+(\uu)/\s_+$, \ie essentially $\n(\uu)/\n_+$ will
fluctuate taking values sensibly different from their average value $1$,
at very rare intervals of time: but when such fluctuations will occurr
one shall see ``bursts'' of anomalous behavior: \ie the motion will be
``{\it intermittent}'' as in the case discussed in non equilibrium
statistical mechanics.
\*
\1
{\bf 11. Entropy creation rate and entropy driven intermittency.}
\numsec=11\numfor=1\*

Of course if $R$ is large the number of degrees of freedom is large
and intermittency on the scale of the fluid container will not be
observable due to its extreme unlikelyhood (expected and
quantitatively predicted by the fluctuation theorem).

Therefore we look also here, in fluid motions, for a {\it local
fluctuation relation}. Fluids seem particularly suitable for verifying
such local fluctuations relations because dissipation occurs {\it
homogeneously}, \ie friction strength is translation invariant.

This implies that we can regard a very small volume $V_0$ of the fluid
as a system in itself (as always done in the derivation of the basic
fluid equations, \eg see [Ga97b]) and we can expect that the phase
space contraction due to such volume elements is simply $\s(\uu)$,
given by \equ(10.9) or \equ(10.12) (``equivalently'' because of our
equivalence conjectures) with the integrals in the numerator and
denominator being extended to the volume $V_0$ rather than to the
whole box, and expressing (essentially by definition) the ``local
phase space contraction'' $\s_{V_0}(\uu)$.

Then $p=\t^{-1}\ig_{-\t/2}^{\t/2}\s_{V_0}(S_t\uu)/\media{\s_{V_0}}_+$
will have a rate function $\z(p)$ which will verify, under the same
assumptions as in \equ(10.10), a large deviation relation as

$$\z(-p)=\z(p)- p\media{\s_{V_0}}_+\,\th\Eq(11.1)$$
for some $\th$: as mentioned the theoretical value of this slope $\th$
seems currently inaccessible to theory (as the theory proposed in
[BG67], [Ga97a] may need substantial modifications, \cfr comment
following \equ(10.10)). The $\media{\s_{V_0}}$ and $\z(p)$ will be
proportional to $V_0$: $\z(p)=V_0\,\,\lis\z(p)$ with a
$V_0$--independent $\lis\z(p)$. Note that $\z(p)$ depends also on $R$.
\*

{\it The small volume element of the fluid will therefore be subject
to rather frequent variations: in spite of $\z(p)$ being proportional
to $V_0$, because now $V_0$ is not large. The consequent intermittency
phenomena can therefore be observed. And as in \S9 once the phase
space contraction is intermittent all properties of the system show
the same behavior.}
\*

And in fact intermittency in observations averaged ove a time span
$\t$ will appear with a time frequency of the form
$e^{V_0\,(\lis\z(p)-\lis\z(1))\,\t}$: the quantity $p$ can be
interpreted as a measure of the ``{\it strength of intermittency}''
observable in easurements averaged over a time $\t$ because as noted
in \S9 and in [Ga99b] the size of $p$ controls the statistical
properties of ``most'' other observables. {\it Therefore the function
$\lis\z(p)$ (hence $\z(p)$) might be directly measurable and it should
be rather directly related to the quantities that one actually
observes in intermittency experiments.} And the difference
$\z(p)-\z(-p)$ can be tested for linearity in $p$ as predicted by the
analysis above.
\*

Note that in an extended system the volume $V$ is much larger than
$V_0$ and we shall see ``for sure'' intermittency (for observables
averaged over a time $\t$) of strength $p$ in a region of volume $V_0$
{\it somewhere} within a volume $W$ such that

$$\fra{W}{V_0} \, e^{V_0\,(\lis \z(p)-\lis\z(1))\t}\simeq 1\Eq(11.2)$$
\*

At this point it seems relevant to recall that it is rather heatedly
being debated whether the name of ``{\it entropy creation rate}'' that
some authors (including the present one) give to the phase space
contraction rate is justified or not, see [An82]. The above properties
not only propose the physical meaning of the quantity $p$ and bring up
the possibility of measuring its rate function $\z(p)$ in actual
experiments but also provide a further justification of the name given
to $\s$ as ``entropy creation rate'' and fuel for the debate that {\it
inevitably} the word entropy generates at each and every occurrence.
\*
\1

{\bf\S12. Benard convection, intermittency and the
Ciliberto--Laroche experiment.}
\numsec=12\numfor=1\*

A very interesting attempt at checking some of the above ideas has been
made recently by Cilberto and Laroche in an experiment on real fluids
which has been performed with the aim of testing the relation \equ(11.1)
locally in a small volume element, [CL98].  By ``real'' we mean here
{\it non numerical}: a distinction that, however, has faded away
together with the XX--th century but that some still cherish: the system
is physically macroscopic (water in a container of a size of the order
of a liter).

This being a real experiment one has to stretch quite a bit the very
primitive theory developed so far in order to interpret it and one has
to add to the chaotic hypothesis other assumptions that have been
discussed in [BG97], [Ga97a] in order to obtain the fluctuation
relation \equ(10.12) and its local couterpart \equ(11.1).

The experiment attempts at measuring a quantity that is eventually
interpreted as the difference $\z(p)-\z(-p)$, by observing the
fluctuations of the product $\th u^z$ where $\th$ is the {\it deviation}
of the temperature from the average temperature in a {\it small volume
element} $\D$ of water at a fixed position in a Couette flow and $u^z$
is the velocity in the $z$ direction of the water in the same volume
element.

The result of the experiment is in a way quite unexpected: it is found
that the function $\z(p)$ is rather irregular and lacking symmetry
around $p=1$: {\it nevertheless the function} $\z(p)-\z(-p)$ {\it seems
to be strikingly linear}. As discussed in [Ga97a], predicting the slope
of the entropy creation rate would be difficult but if the equivalence
conjecture considered above and discussed more in detail in [Ga97a] is
correct then we should expect linearity of $\z(p)-\z(-p)$.

In the experiment of [CL98] the quantity $\th u_z$ did not appear to be
the divergence of the phase space volume simply because there was no
model proposed for a theory of the experiment. Nevertheless
Ciliberto--Laroche select the quantity $\int_\D \th\,u^z \,d\V x$ on the
basis of considerations on entropy and dissipation so that there is
hope that in a model of the flow this quantity can be related to
the entropy creation rate discussed in \S10,\S11.

Here we propose that a model for the fluid, that can be reasonably
used, is Rayleigh's model of convection, [Lo63], [LL71] and [Ga97b]
sec. 5. An attempt for a theory of the experiment could be the
following.

One supposes that the equations of motion of the system in the {\it
whole container} (of linear size of the order of $30\,cm$) are written
for the quantities $t,x,z,\th,\uu$ in terms of the height $H$ of the
container (assumed to be a horizontal infinite layer), of the
temperature difference between top and bottom $\d T$ and in terms of the
phenomenological ``{\it friction constants}'' $\n,\ch$ of viscosity,
dynamical thermal conductivity and of the thermodynamic dilatation
coefficient $\a$. We suppose that the fluid is $3$--dimensional but
stratified, so that velocity and temperature fields do not depend on the
coordinate $y$, and gravity is directed along the $z$--axis: $\V g= g\,
\V e,\, \V e=(0,0,-1)$. The temperature deviation $\th$ is defined as
the difference betwen the temperature $T(x,y,z)$ and the temperature
that the fluid would have at height $z$ in absence of convection, \ie
$T_0-z\,\d T/H$ if $T_0$ is the bottom temperature.

In such conditions the equations, including the boundary conditions
(of fixed temperature at top and bottom and zero normal velocity at
top and bottom), the convection equations in the Rayleigh model, see
[Lo63] eq. (17), (18) where they are called the {\it Saltzman
equations}, and [Ga97b] \S1.5, become
$$\eqalignno{
&\V\dpr\cdot\V u=0,\kern5cm \int u_x d\V{x}=\int u_yd\V{x}=0\cr
&\dot{\V u}+\W u\cdot\W \partial\,\V u=\n\D\V
u-\a\th\V g-\V\dpr p',&\eq(12.1)
\cr
&\dot\th+\W
u\cdot\W\partial\th=\chi\D\th+{\d T\over H} u_z
\qquad \th(0)=0=\th(H),\quad
u_z(0)=0=u_z(H),%\cr &\int u_x d\V{x}=\int u_yd\V{x}=0\cr}$$
\cr}$$
The function $p'$ is related to the pressure $p$: within the
approximations it is $p=p_0-\r_0 g z+ p'$. We shall impose for
simplicity horizontal periodic boundary conditions in $x, y$ so that
the fluid can be considered in a finite container $ V$ of side $a$ for
some $a>0$ prefixed (which in theoriginal variable would correspond to
a container of horizntal size $a H$).

It is useful to define the following adimensional quantities
$$\eqalign{
\t&=t\n H^{-2},\ \x= x H^{-1},\ \h=y H^{-1},\ \z=z H^{-1},\cr
\th^0&=\fra{\a\th}{\a\,\d T},\ \V u^0=(\sqrt{g H\a\,\d T})^{-1}\,\V
u\cr
R^2&={g H^3\a\,\d T\over\n^2},\qquad R_{Pr}={\n\over\k}\cr}
\Eq(12.2)$$
and one checks that the Rayleigh equations take the form
$$\eqalign{
&\dot\V u+ R\W u\cdot\W\partial\,\V u=\D\V u- R\th\V e-\V\dpr p,\cr
&\dot\th+ R\W u\cdot\W\partial\,\th=R_{Pr}^{-1}\D\th+ R u_z,\cr
&\V\dpr\cdot\V u=0\cr
&u_z(0)=u_z(1)=0,\qquad\th(0)=\th(1)=0,\cr
&\ig_V u_x d\V{x}=\ig_V u_y
d\V{x}=0\cr}\Eq(12.3)$$
where we again call $t,x,y,z,\V u,\th$ the adimensional coordinates
$\t,\x,\h,\z, \V u^0,$ $\th^0$ in \equ(12.2). The numbers $R, R_{Pr}$ are
respectively called the {\it Reynolds and Prandtl numbers} of the
problem: $R_{Pr}=\sim6.7$ for water while $R$ is a parameter that we
can adjust, to some extent, from $0$ up to a rather large value.

According to the principle of equivalence stated in [Ga97a] here one
could impose the constraints

$$
\int_V \big( \,\V u^2\,+\,\fra1{R_{Pr}}
\th^2\,\big)\,d\V x= C\Eq(12.4)$$
on the ``{\it frictionless equations}'' (\ie the ones without the
terms with the laplacians) and determine the necessary forces via
Gauss' principle of minimal effort, see footnote ${}^3$ and [Ga96a],
[Ga97a]. We use as {\it effort functional} of an acceleration field
$\V a$ and of a temperature variation field $s$ the quantity
$$\eqalignno{
&\EE(\V a,s)\defi \big((\V a+\V\dpr p-\V f),(-\D)^{-1}(\V a+\V\dpr p-\V
f)\big)+&
\eq(12.5)\cr
&+\big((s-\f),(-\D)^{-1}(s-\f)\big)\kern2cm {\rm with}\cr
&\V f\defi -R \th \V e,\qquad \f \defi R u_z\cr}$$
and require it to be minimal over the variations $\V\d(\V x)$ of $\V
a=\fra{d \V{ u}}{dt}$ and $\t(\V x)$ of $s=\fra{d \th}{dt}$ with the
constraints that for all $\V x$ it is $\V\partial \cdot \V \d=0$,
besides those due to the boundary conditions. The result is

$$\eqalign{
&\V\dpr\cdot\V u=0\cr
&\dot{\V u}+R\,\W u\cdot\W \partial\,\V u=R\th\,\V e-\V\dpr p'+
\V\t_{th}
\cr
&\dot\th+R \W u\cdot\W\partial\th=Ru^z+\l_{th}\cr
&\th(0)=0=\th(H),\quad \int_V u_x d\V{x}=\int_V u_yd\V{x}=0\cr}
\Eq(12.6)$$
where the frictionless equations are modified by {\it the thermostats
forces} $\V\t_{th},\l_{th}$: the latter impose the nonholonomic
constraint in \equ(12.4) with the effort functional defined by
\equ(12.5). Looking only at the bulk terms we see that the equations
obtained by imposing the con1straints via Gauss' principle become the
\equ(12.3) with coefficients in front of the Laplace operators equal
to $\n_G,\n_G R_{Pr}^{-1}$, respectively, with the ``gaussian
multiplier'' $\n_G$ being an {\it odd} functions of $\V u$, see
[Ga97a]: setting ${\tilde {C}}_V (\V u,\th)=\int_V
\big((\W\partial \V u)^2+R_{Pr}^{-1}(\V\partial \th)^2\big) d\V x$
one finds
$$
\nu_G={\tilde C}_V
(\V u,\th)^{-1}R(1+R_{Pr}^{-1})\int_V
\,u^z\th\,\,d\V x\Eq(12.7)$$
And the equations become, finally

$$\eqalign{ &\V\dpr\cdot\V u=0\cr &\dot{\V u}+R\,\W u\cdot\W
\partial\,\V u=R\th\,\V e-\V\dpr p'+\n_G \,\D\V u\cr &\dot\th+R \W
u\cdot\W\partial\th=Ru^z+\n_G\fra1{R_{Pr}}\,\D \th\cr
&\th(0)=0=\th(H),\quad \int u_x d\V{x}=\int u_yd\V{x}=0\cr}
\Eq(12.8)$$
If one wants the equivalence between the ensembles of SRB
distributions for the equation \equ(12.8) and for \equ(12.3) one
has to tune, [Ga97a], the value of the constant $C$ in
\equ(12.4) so that the time average value $\media{\n_G}_+$ of $\n_G$
is precisely the physical one: namely $\media{\n_G}=1$ by
\equ(12.3). This is (again) the same, in spirit, as fixing the
temperature in the canonical ensemble so that it agrees with the
microcanonical temperature thus implying that the two ensembles give the
same averages to the local observables.
\*

The equations \equ(12.8) are time reversible (unlike the \equ(12.3))
under the time reversal map:

$$
(\V u,\th)=(-\V u,\th)\Eq(12.9)$$
and they should be supposed, by the arguments in [Ga97a] and \S10,11:
``equivalent'' to the irreversible ones \equ(12.3),

The \equ(12.8) should therefore have a ``divergence'' $\s(\V u,\th)$
whose fluctuation function $\z(p)$ verifies a linear fluctuation
relation, \ie $\z(p)-\z(-p)$ should be linear in $p$. Note that the
divergence of the above equations is proportional to $\n_G$ if one
supposes that the high momenta modes with $|\kk|>K(R)=R^\k$ with $\k$
suitable can be set equal to $0$ so that the equation
\equ(12.8) becomes a system of finite differential equations for
the Fourier components of $\V u,\th$.

For instance the Lorenz' equations, [Lo63] see also \S17 of [Ga97b],
reduced the number of Fourier components necessary to describe
\equ(12.3) to just three components, thus turning it into a system of
three differential equations.

Proceeding in this way the divergence of the equations of motion can
be computed as a sum of two integrals one of which proportional to
$\n_G$ in \equ(12.7).  If instead of integrating over the whole sample
we integrate over a small region $\D$, like in the experiment of
[CL98], we can expect to see a fluctuation relation for the entropy
creation rate if the fluctuation theorem {\it holds locally}, \ie
for the entropy creation in a small region.

As for the cases in \S11 this is certainly not implied by the proof in
[GC95]: however {\it when the dissipation is homogeneous through the
system}, as it is the case in the Rayleigh model there is hope that
the fluctuation relation holds locally because ``a small subsystem
should be equivalent to a large one''. As noted in \S9 the actual
possibility of a local fluctuation theorem in systems with homogeneous
dissipation has been shown in [Ga99c], after having been found through
numerical simulations in [GP99], and this example was relevant because
it gave us some justification to imagine that it might apply to the
present situation as well.

The entropy creation is due to the term $R\int_\D u_z \th\,d\V x/
{\tilde C}_\D(\V u,\th)$, where
$\D$ is the region where the measurements of [CL98] are performed,
hence we have a proposal for the explanation of the remarkable
experimental result. Unfortunately in the experiment [CL98] the
contributions not explicitly proportional to $R$ to the entropy
creation rates have not been measured nor has been the $\hat C$ in
\equ(12.7) which also fluctuates (or might fluctuate). In any event
they might be measurable by improving the same apparatus, so that one
can check whether the above attempt to an explanation of the
experiment is correct, or try to find out more about the theory in
case it is not right. If correct the above ``theory'' the experiment
in [CL98] would be quite important for the status of the chaotic
hypothesis.
\*

{\bf\S13. Conclusions.}
\*

The chaotic hypothesis promises a point of view on non equilibrium
that has proved so far of some interest. Here we have exposed the
basic ideas and attempted at drawing some consequences: admittedly the
most interesting rely on rather phenomenological and heuristic
grounds. They are summarized below.
\*
(1) The definition of nonequilibrium ensembles with the proposal that
    out of equilibrium also the equation of motion should be
    considered as part of the definition of ensemble. This is take
    into account that while in equilibrium the system is uniquely
    defined by its microscopic forces and constituents {\it in non
    equilibrium it is not so}. Systems must be put in contact with
    thermostats if we want them to become stationary after a transient
    time. And (for large systems) there may be several equivalent ways
    of taking heat out of a system, \ie several thermostats, without
    affecting the properties of stationary state that is eventually
    reached by the system itself.

(2) Equivalence of ensembles has the most striking aspect that systems
    which evolve with equations that are very different may exhibit
    the same statistical properties. In particular reversible
    evolutions might be equivalent to non reversible ones, thus making
    it possible to apply results that require reversibility, in
    particular the fluctuation relations, to cases in which it is not
    valid.

(3) An interpretation of the quantity $p$ that intervenes in the
    fluctuation theorems in terms of an intermittency phenomenon and as
    a further quantitative measure of it.

(4) The possibility of applying the theory to strongly turbulent
    motions was the origin of the Ruelle's principle that evolved into
    the chaotic hypothesis: therefore not surprisingly the ideas can
    be applied to fluid dynamics. We have discussed a possible
    approach. The approach leads again to a proposal for the theory of
    certain intermittency phenomena which appear quantitatively
    related to entropy creation fluctuations.

(5) The possibility of measurement of the rate $\z(p)$ leads to a a
    possible prediction of the spatial frequency of internittent
    events of strength $p$ or, as I prefer, with entropy creation rate
    $p$ (see (4) above, \equ(11.2) and \S12). This seems testable in
    concrete experiments (both real and numerical).

(6) We have used the results in (2)\%(8) to hint at an
    interpretation of the experiment by Ciliberto and Laroche on Benard
    convection in water.
\*

Although the theory is still at its beginning and it migh turn out to
be not really of interest it seems that at this moment it is worth
trying to test it both in its safest, \cfr \S2\%\S8, and in its most
daring, \cfr \S9\%\S12, predictions.

\*
{\bf Acknowledgements: \it I am greatly indebted to Professor
R. Newton for giving me the opportunity to collect the above thoughts
which, collected together, go far beyond what I originally planned
after he proposed to me to write this review. Work partially supported
by Rutgers University and by MPI through grant \# ??????????.}
\*

\1
{\bf References.}
\*

\def\*{\vskip2pt}\def\0{\*\noindent}
\0[An82] Andrej, L.: {\it The rate of entropy change in
non--Hamiltonian systems}, Physics Letters, {\bf 111A}, 45--46, 1982.
And {\it Ideal gas in empty space}, Nuovo Cimento, {\bf B69}, 136--144,
1982. See also {\it The relation between entropy production and
$K$--entropy}, Progress in Theoretical Physics, {\bf75}, 1258--1260,
1986.

\0[Be50] Becker, R.: {\sl Teoria dell' elettricit\`a}, Sansoni, vol. II, 1950

\0[Bo66] Boltzmann, L.: {\it\"Uber die mechanische Bedeutung des
zweiten Haupsatzes der W\"armetheorie}, in
"Wissenschaftliche Abhandlungen", ed. F. Hasen\"ohrl,
vol. I, p. 9--33, reprinted by Chelsea, New York.

\0[Bo84] Boltzmann, L.: {\it \"Uber die Eigenshaften monzyklischer
und anderer damit verwandter Systeme}, in "Wissenshafltliche
Abhandlungen", ed. F.P. Hasen\"ohrl, vol. III, p. 122--152,
Chelsea, New York, 1968, (reprint).

\0[BGG97] Bonetto, F., Gallavotti, G., Garrido, P.: {\it
Chaotic principle: an experimental test}, Physica D, {\bf 105},
226--252, 1997.

\0[BG97] Bonetto, F., Gallavotti, G.: {\it Reversibility, coarse graining
   and the chaoticity principle}, Communications in Mathematical
   Physics, {\bf189}, 263--276, 1997.

\0[CG99] Cohen, E.G.D., Gallavotti, G.: {\it Note on Two Theorems in
   Nonequilibrium Statistical Mechanics}, Journal of Statistical
   Physics, {\bf 96}, 1343--1349, 1999.

\0[CL98] Ciliberto, S., Laroche, C.: {\it An experimental
verification of the Galla\-vot\-ti--Cohen fluctuation theorem}, Journal de
Physique, {\bf8}, 215--222, 1998.

\0[ECM93] Evans, D.J.,Cohen, E.G.D., Morriss, G.P.: {\it Probability
of second law violations in shearing steady f\/lows}, Physical Review
Letters, {\bf 71}, 2401--2404, 1993.

\0[EM90] Evans, D.J., Morriss, G.P.: {\sl Statistical
Mechanics of Non\-equilibrium fluids}, Academic Press, New York, 1990.

\0[GC95] Gallavotti, G., Cohen, E.G.D.: {\it Dynamical ensembles in
   nonequilibrium statistical mechanics}, Physical Review Letters,
   {\bf74}, 2694--2697, 1995. And: {\it Dynamical ensembles in
   stationary states}, Journal of Statistical Physics, {\bf 80},
   931--970, 1996].

\0[Ga95] Gallavotti, G.: {\it Topics on chaotic dynamics}, in Third
  Granada Lectures in {\sl Computational Physics}, Ed. P. Garrido,
  J. Marro, in Lecture Notes in Physics, Springer Verlag, {\bf 448},
  p. 271--311, 1995.

\0[Ga96a] Gallavotti, G.: {\it Extension of Onsager's reciprocity to
   large fields and the chaotic hypothesis}, Physical Review Letters,
   {\bf 77}, 4334--4337, 1996.

\0[Ga96c] Gallavotti, G.: {\it New methods in nonequilibrium gases and
fluids}, Open Systems and Information Dynamics, Vol. 6, 101--136, 1999
(original in chao-dyn \#9610018).

\0[Ga96b] Gallavotti, G., Ruelle D.: {\it SRB states and nonequilibrium
   statistical mechanics close to equilibrium}, Communications in
   Mathematical Physics, {\bf190}, 279--285, 1997.

\0[Ga97a] Gallavotti, G.: {\it Dynamical ensembles equivalence in fluid
   mechanics}, Physica D, {\bf 105}, 163--184, 1997.

\0[Ga97b] Gallavotti, G.: {\it Ipotesi per una introduzione alla Meccanica
  dei Fluidi}, ``Quader\-ni del CNR-GNFM'', vol. {\bf 52}, p. 1--428,
  Firenze, 1997. English translation in progress: available at
  http:$\backslash\backslash$ipparco.roma1.infn.it.

\0[Ga98] Gallavotti, G.: {\it Chaotic dynamics, fluctuations,
   non-equilibrium ensembles},\\ Chaos, {\bf8}, 384--392, 1998. See also
   [Ga96c].

\0[Ga99a] Gallavotti, G.: {\it Statistical Mechanics}, Springer Verlag,
1999.

\0[Ga99b] Gallavotti, G.: {\it Fluctuation patterns and conditional
  reversibility in nonequilibrium systems}, Annales de l' Institut
  H. Poincar\'e, {\bf70}, 429--443, 1999.

\0[Ga99c] Gallavotti, G.: {\it A local fluctuation theorem}, Physica A,
{\bf 263}, 39--50, 1999. And {\it Chaotic Hypothesis and Universal Large
Deviations Properties}, Documenta Mathematica, extra volume ICM98,
vol. I, p. 205--233, 1998, also in chao-dyn 9808004.

\0[Ge97] Gentile, G.: {\it Large deviation rule for Anosov flows}, 
Forum Mathematicum, {\bf10}, 89--118, 1998.

\0[GP99] Gallavotti, G., Perroni, F.: {\it An experimental test of the
local fluctuation theorem in chains of weakly interacting Anosov
systems}, preprint, 1999, in http://ipparco. roma1. infn. it at the 1999
page.

\0[HHP87] Holian, B.L., Hoover, W.G., Posch. H.A.:
{\it Resolution of Loschmidt's paradox: the origin of irreversible
behavior in reversible atomistic dynamics}, Physical Review Letters,
{\bf 59}, 10--13, 1987.

\0[Lo63]  Lorenz, E.: {\it Deterministic non periodic flow}, J.
of the Athmospheric Sciences, 20, 130- 141, 1963.

\0[LL71] Landau, L., Lifchitz, E.: {\sl M\'ecanique des fluides}, MIR,
Moscou, 1971.

\0[RS99] Rondoni, L., Segre, E.: {\it Fluctuations in two dimensional
reversibly damped turbulence}, Nonlinearity, {\bf12}, 1471--1487, 1999.

\0[Ru76] Ruelle, D.: {\it A measure associated with Axiom A
attractors}, American Journal of Mathematics, {\bf98}, 619--654, 1976.

\0[Ru78] Ruelle, D.: {\it Sensitive dependence on initial conditions
and turbulent behavior of dynamical systems}, Annals of the New York
Academy of Sciences, {\bf356}, 408--416, 1978. This is the first place
where the hypothesis analogous to the later chaotic hypothesis was
formulated (for fluids): however the idea was exposed orally at least
since the talks given to illustrate the technical work [Ru76], which
appeared as a preprint and was submitted for publication in 1973 but was
in print three years later.

\0[Ru96] Ruelle, D.: {\it Positivity of entropy production in
nonequilibrium statistical mechanics}, Journal of Statistical Physics,
{\bf 85}, 1--25, 1996. And Ruelle, D.: {\it Entropy production in
nonequilibrium statistical mechanics}, Communications in Mathematical
Physics, {\bf189}, 365--371, 1997.

\0[Ru99a] Ruelle, D.: {\it Smooth dynamics and 
new theoretical ideas in non-equilibrium statistical mechanics},
Journal of Statistical Physics, {\bf 95}, 393--468, 1999.

\0[Ru99b] Ruelle, D.: {\it A remark on the equivalence of isokinetic
and isoenergetic thermostats in the thermodynamic limit}, preprint
IHES 1999, to appear in Journal of Statistical Physics.

\0[Se87] Seitz, F.: {\sl The modern theory of solids}, Dover, 1987
(reprint).
\vskip0.5cm
\def\*{\vskip3pt}

\FINE

\ciao